\documentclass[11pt,a4paper]{article}
\usepackage{jcappub}
\usepackage{epsfig}
\input amssym.def
\input amssym.tex
\usepackage{epsfig}
\usepackage{graphicx}
\usepackage{dcolumn}
\usepackage{bm}
\usepackage{color}
\newcommand{\be}{\begin{equation}}
\newcommand{\ee}{\end{equation}}
\newcommand{\bk}{\mbox{\boldmath$k$}}
\newcommand{\bx}{\mbox{\boldmath$x$}}
\newcommand{\scbk}{{\scriptsize\mbox{\boldmath$k$}}}
\newcommand{\scbx}{{\scriptsize\mbox{\boldmath$x$}}}
\renewcommand{\[}{\begin{equation}}
\renewcommand{\]}{\end{equation}}
\renewcommand{\]}{\end{equation}}

\def\thebiblio#1{
\begin{center}\bf \large References
\end{center}
\list
{[\arabic{enumi}]}{\settowidth\labelwidth{#1.}\leftmargin\labelwidth
 \advance\leftmargin\labelsep
 \usecounter{enumi}}
 \def\newblock{\hskip .11em plus .33em minus -.07em}
 \sloppy
 \sfcode`\.=1000\relax}

\subheader{\small HIP-2012-26/TH}
\title{On the fate of coupled flat directions during inflation}

\author[a]{Juan C. Bueno S\'anchez,}
\author[b]{Kari Enqvist}%
\affiliation[a]{Departamento de F\'isica At\'omica, Molecular y Nuclear, Universidad Complutense de Madrid,\\
28040 Madrid, Spain}
\affiliation[b]{University of Helsinki and Helsinki Institute of Physics,\\
FIN-00014, University of Helsinki, Finland}

\abstract{We investigate the stochastic dynamics of the long wavelength modes of a generic light scalar field that during inflation is coupled to another scalar field. The coupling plays an important role for the fluctuation of the field amplitude and may block its initial growth. We find that such a blocking is avoided, albeit only temporarily, if the light scalar has an initial non-vanishing expectation value $\langle\phi\rangle$ larger than a certain critical value, for which we provide an estimate. We also show that the field fluctuations will eventually reach an equilibrium amplitude provided inflation is sufficiently long-lasting. We present a novel, general expression for the variance  $\langle\phi^2\rangle$ that takes into account the coupling of the massless field and describes the growth during the epoch of quasi-free fluctuations as well as the late-time approach to the equilibrium.}

\begin{document}
\maketitle

\section{Introduction}

During inflation, all light scalar fields are subject to stochastic fluctuations  \cite{Linde:1982uu,Starobinsky:1982ee,Vilenkin:1982wt}. Their amplitudes tend to increase by a process that could be described as random walk. Once inflation is over, the fields can start moving classically, with many interesting cosmological consequences. An example is Affleck-Dine baryogenesis \cite{Affleck:1984fy}, where the motion of a complex scalar field, initially displaced from the minimum of its potential by inflationary fluctuations, dynamically generates a baryon asymmetry. However, if there are several scalars that are coupled to each other, their dynamics during inflation can be rather complicated as the fluctuation of one scalar can backreact on the fluctuations of another. This is also the situation one would encounter in theories with flat directions  (or moduli fields), which are rays in field space along which the potential either vanishes completely or is very small. An example is the
Minimal Supersymmetric Standard Model (MSSM), which has some three hundred flat directions; these have all been classified (for a review, see \cite{MSSM-REV}). In the MSSM flat directions correspond to configurations where some of the field values are related to each other while the rest are set to zero. As a consequence, most of the flat directions are not simultaneously flat, and if there is a large field fluctuation along some given flat direction, many other potentially flat directions will no longer remain flat. In effect, fluctuations may provide some of the flat directions with effective masses that prevent the growth of the amplitudes, as was discussed in \cite{Enqvist:2011pt}.

The dynamics of such a system of coupled, stochastically fluctuating scalar fields is in general very complicated. In the present paper we attempt to give a careful analysis of the growth of fluctuations in a system that only has two scalars, coupled to each other. Unlike in \cite{Enqvist:2011pt}, we do not require that the initial field (expectation) values are zero but allow also for the possibility that at the onset of inflation, the light field (flat direction) can have an arbitrary expectation value. We assume that inflation is independent of these scalars, which for the purposes of inflationary dynamics are just spectators, and that the Hubble rate $H$ can be taken as a constant. We further assume an FRW universe with $H=\dot a/a$, where $a(t)$ is the scale factor, and the number of $e$-foldings given is by $N=\ln(a(t)/a(0))$. By definition, for a given background value $\phi_0$, the field $\phi$ is called light if $V''(\phi_0)\ll H^2$.

The paper is organized as follows. In section 2, the basics of the stochastic approach to inflation are reviewed. In section 3, we study the behavior of a massless scalar field coupled to a massive one, and approximate the behavior of the massless field when the former begins with a large expectation value. In section 4 we investigate the consequences of such an assumption and derive an expression for the inflationary fluctuations of the coupled massless field. In section 5 we present our conclusions.

\section{Stochastic formalism}
In general, the equation of motion for a minimally coupled, self-interacting scalar field with mass $M$ and interaction potential $V(\phi)$ is
\[
\ddot\Phi+3H\dot\Phi+M^2\Phi^2-a^{-2}\nabla^2\Phi+V'(\Phi)=0\,.
\]
Provided $M^2\ll H^2$, the field $\Phi$ undergoes quantum particle production from its vacuum fluctuation during inflation. Owing to the probabilistic nature of the vacuum fluctuation, the process of particle production is stochastic. Also, due to the inflationary expansion, the momenta of the created quanta are redshifted as the corresponding field modes become superhorizon. Consequently, the evolution of the light field $\Phi$ on long wavelength scales contains an stochastic component due to the inflationary particle production.

In order to study the stochastic evolution of a light field $\Phi$ during inflation, one introduces a coarse-graining scale \mbox{$k_s(t)\equiv\epsilon a(t) H$}, where $\epsilon\ll1$ \cite{Starobinsky:1986fx}. This scale serves as a divide to decompose the field $\Phi$ into its long and short wavelength parts as follows
\[\label{eq10}
\Phi(t,\bx)=\phi(t,\bx)+\phi_q(t,\bx)\,,
\]
where the short wavelength part of the field is given by
\[
\phi_q(t,\bx)=\int\frac{d^3\bk}{(2\pi)^3}\,\theta(k-k_s)\left[\hat a_\scbk\varphi_k(t)\,e^{i\scbk\cdot\scbx}+\hat a_\scbk^\dagger\varphi_k^*(t)e^{-i\scbk\cdot\scbx}\right]\,,
\]
where $k=|\bk|$, describes the quantum behavior of the scalar field on subhorizon scales.  The long wavelength part $\phi$ (left unspecified), describes the field's classical behavior on superhorizon scales.

The equation of motion of a scalar field in de Sitter space can be written as \cite{Hosoya:1988yz}
\[\label{eq15}
\ddot\phi+3H\dot\phi-a^{-2}\nabla^2\phi+V'(\phi)=\xi(x,t)\,,
\]
where
\[
\xi(x,t)=-\left(\ddot\phi_q+3H\dot\phi_q-a^{-2}\nabla^2\phi_q+V'(\phi_q)\right)
\]
is a stochastic source of white source, i.e. $\langle\xi(x,t)\rangle=0$, where $\langle\,\rangle$ stands for the vacuum expectation value of quantum operators.

To simplify the discussion, we adopt the so-called separate Universe approach \cite{Wands:2000dp}, according to which the Universe can be regarded as an ensemble of patches of size $k_s^{-1}$ in which the field is approximately homogeneous and evolves independently of the other patches. Under this assumption, the term $\nabla^2\phi$ can be neglected in Eq.~(\ref{eq15}) and the stochastic source $\xi$ becomes space-independent within each coarse-graining volume. In the case of minimally coupled, light scalar fields, the self-correlation function of the stochastic source is
\[\label{eq40}
\langle\xi(t)\xi(t')\rangle=\frac{9H^5}{4\pi^2}\,\delta(t-t')\,.
\]
Regarding the above average $\langle\,\rangle$, we wish to emphasize that for a free scalar field with mass $m\ll H$, the quantum average $\langle\phi^2\rangle$ computed by field theoretical methods \cite{Bunch:1978yq} agrees with the average over an ensemble of independent universes \cite{Mijic:1994vv}. Therefore, when applied to a light scalar field $\phi$ during inflation, we are entitled to interpret the quantum average $\langle\phi^2\rangle$ as an ensemble average over coarse-graining patches.

We want to address the growth of field fluctuations in the regime in which $\phi$ performs a slow-roll motion due to the flatness of its potential. In such case, the solution to Eq.~(\ref{eq15}) can be approximated by neglecting the second time derivative $\ddot \phi$. The behavior of the coarse-grained field is then determined by the Langevin equation
\[\label{eq11}
3H\dot\phi+V'(\phi)\simeq\xi(t)\,.
\]
From the standpoint of stochastic equations, it is straightforward to write down the Fokker-Planck equation \cite{Gardiner}
\[\label{eq4}
\partial_tP(\phi,t)=\partial_\phi\left(\frac{V'(\phi)}{3H}\,P(\phi,t)\right)+\frac{H^3}{8\pi^2}
\,\partial_{\phi\phi}^2
P(\phi,t)
\]
describing the evolution of the probability density that the field $\Phi$ has an expectation value $\phi$ on a separate Universe, or coarse-graining patch. The first term on the righthand side accounts for the deterministic motion of the field, or drift term, and the second for its random motion.

In the simple case of a free scalar field with mass $m$ and the potential $V(\phi)=\frac12\,m^2\phi^2$, whenever the field is displaced from the origin, the restoring force driving the field to $\phi=0$ increases proportionally to $\phi$. For sufficiently large $\phi$, the restoring force compensates for the diffusive motion of the field and an equilibrium situation arises. Since in the equilibrium the probability distribution becomes stationary with $\partial_tP(\phi,t)=0$, demanding that $P(\phi,t)$ vanishes sufficiently fast at $\phi\to\pm\infty$ one readily obtains the well-known equilibrium amplitude of the field fluctuations \cite{Starobinsky:1994bd}
\[\label{eq22}
\langle\phi^2\rangle_{\rm eq}=\frac{3H^4}{8\pi^2m^2}\,.
\]

In relation to this result, in Ref.~\cite{Enqvist:2011pt} it is shown that flat directions of the MSSM reach a stationary fluctuation amplitude similar to the above when they begin with a vanishing expectation value. This results arises because the coupling of the flat direction to other fields provides the former with an effective mass, which then results in the emergence of an stationary distribution. As we shall show in the following sections, if the coupled massless field (or the would-be flat direction) begins with a large expectation value $\langle\phi\rangle\gg H$, the amplitude of the field fluctuations reaches the equilibrium value only after a sufficiently long wait. However, the stochastic dynamics in the coupled massless field in this case differs substantially from the one studied in \cite{Enqvist:2011pt}. Also, and in agreement with recent numerical results \cite{Kawasaki:2012bk}, we find that the coupled massless field is able to fluctuate as if free. Nevertheless, in our approach we discover that this regime of quasi-free fluctuations applies for a limited time only. Beyond certain timescale, the system engages into a lengthy process ending up with the massless field fluctuating with its equilibrium amplitude in the entire ensemble of coarse-graining patches.

\section{Behavior in the large field limit}
In this section we investigate a system of two scalar fields, denoted by $\phi$ and $\chi$, minimally coupled to gravity during inflation. We assume that the energy density of both fields remains always subdominant, hence their evolution does not affect the inflationary expansion. Assuming that $\phi$ is massless while $\chi$ has a squared mass $\bar m_\chi^2$, the Lagrangian of the system is
\[\label{eq16}
{\cal L}=\frac12\partial_\mu\Phi\partial^\mu\Phi+\frac12\partial_\mu\chi\partial^\mu\chi
-\frac12\bar m_\chi^2\chi^2-V(\Phi,\chi)\,,
\]
where the interaction potential is
\[\label{eq13}
V(\Phi,\chi)=\frac12g^2\Phi^2\chi^2~,
\]
and $g$ is a coupling constant. Such an interaction potential is ubiquitous in quantum field theory, and its consequences have been extensively studied in the theory of reheating and preheating \cite{Dolgov:1982th,Abbott:1982hn,Kofman:1994rk,Felder:1998vq}. Moreover, this coupling has been shown to provide a mechanism to trap string moduli fields at points of enhanced symmetry \cite{Kofman:2004yc,Watson:2004aq}, also to result in trapped inflation \cite{Kadota:2003tn,Bueno Sanchez:2006ah,Green:2009ds}, and to give rise to a source of non-gaussianity in the inflaton's perturbation spectrum \cite{Barnaby:2009mc,Wu:2006xp,Lee:2011fj}. Typically, the moduli field playing the role of the inflaton approaches a location in field space where the $\chi$ field becomes massless. If the light field $\Phi$ becomes sufficiently close to the point of enhanced symmetry, a $\chi$-particle production mechanism is triggered which modifies the dynamics of $\Phi$. In this paper we do not study this situation in any significant detail, but will invoke the existence and implications of such a process.

The equation of motion for the fields $\Phi$ and $\chi$ is
\[\label{eq1}
\ddot\Phi+3H\dot\Phi-a^{-2}\nabla^2\Phi+g^2\chi^2\Phi=0~,
\]
\[
\ddot\chi+3H\dot\chi-a^{-2}\nabla^2\chi+\left(\bar m_\chi^2+g^2\Phi^2\right)\chi=0\,.
\]
The effective masses of $\Phi$ and $\chi$ are determined by
\[
m_\Phi^2\equiv g^2\chi^2\quad,\quad m_\chi^2\equiv \bar m_\chi^2+g^2\Phi^2\,.
\]

To solve the dynamics of the long wavelength part of $\Phi$, denoted by $\phi$ (see Eq.~(\ref{eq10})), our starting assumption is that within coarse-graining patches $\Phi$ is approximately homogeneous (hence determined by $\phi$) and that it has an initial non-vanishing expectation value $\phi_0$, i.e.
\[
\Phi(x,0)\approx\phi_0\,,
\]
large enough so that the field $\chi$ is initially heavy, namely
\[
m_\chi(0)^2\simeq g^2\phi_0^2\gg H^2\,.
\]
Later on we consider and discuss the case for which $\bar m_\chi^2\sim g^2\Phi^2$.

\subsection{Integrated coarse-grained dynamics}
Since $\chi$ is a heavy field, it performs oscillations around $\chi=0$ in a timescale that is very short compared to the Hubble time. On the other hand, we wish to consider the situation in which the massless field $\Phi$ receives a small effective mass due to its coupling to $\chi$, and hence performs a slow-roll motion. In such case, $\Phi$ evolves in a timescale large compared to the Hubble time. This timescale separation between the evolutions of $\Phi$ and $\chi$ implies that we can integrate out the field $\chi$ and compute the resulting effective equation of motion for the long wavelength part $\phi$. Neglecting dissipation effects, which we justify later on, the equation for the coarse-grained light field $\phi$ can be written as
\[\label{eq17}
3H\dot\phi+m_\phi^2\phi\simeq\xi(t)\,,
\]
where $m_\phi^2\equiv\langle m_\Phi^2\rangle=g^2\langle\chi^2\rangle$ and we have neglected $\ddot\phi$ and the gradient $a^{-2}\nabla^2\phi$. Of course, to use (\ref{eq17}) consistently we must secure that $\phi$ behaves as a light field, namely that $g^2\langle\chi^2\rangle\ll H^2$. Negligible dissipation effects also imply that we can treat $\xi(t)$ as a stochastic source of gaussian noise whose correlation function is given by Eq.~(\ref{eq40}).

To compute the variance $\langle\chi^2\rangle$ on the coarse-graining scale $k_s$ we need to solve for the mode equation for the $\chi_k$ modes
\[\label{eq32}
\ddot\chi_k+3H\dot\chi_k+\left(m_\chi^2+\frac{k^2}{a^2}\right)\chi_k=0\,.
\]
Owing to the slow-roll motion of $\phi$ we may assume that $m_\chi^2\simeq g^2\phi^2$ does not change substantially from the time when $k^2/a^2=m_\chi^2$ until the $k$-mode exits the coarse-graining scale. In such case, the solution to the mode equation (\ref{eq32}) matching the solution $\chi_k=\frac{a^{-1}}{\sqrt{2k}}\,e^{ik/aH}$ in the limit $k/aH\to\infty$ is approximated by
\[\label{eq19}
\chi_k(t)=a^{-3/2}\sqrt{\frac{\pi}{4H}}\,e^{i\left(\nu+1/2\right)\frac{\pi}{2}}
H_\nu^{(1)}(k/aH)\,,
\]
where $\nu^2\equiv9/4-m_\chi^2/H^2$ and $H_\nu^{(1)}$ is the Hankel function of first kind. Expanding the Hankel function for $k\ll aH$ we obtain
\begin{eqnarray}
|\chi_k|^2&=&\frac{e^{i\pi\nu}\pi}{4a^3H}|H_\nu^{(1)}(k/aH)|^2\nonumber\\
\label{eq38}
&\simeq&\frac{e^{-|\nu|\pi}\pi}{4a^3H}\frac{\left[1+\coth\left(|\nu|\pi\right)\right]^2
\sinh\left(|\nu|\pi\right)}{|\nu|\pi}
=\frac{1+\coth\left(|\nu|\pi\right)}{4a^3H|\nu|}\,.
\end{eqnarray}

In order to compute $\langle\chi^2\rangle$ on the coarse-graining scale we follow the regularization scheme by Vilenkin \cite{Vilenkin:1983xp} (see also \cite{Enqvist:1987au}). The computation of the quantum average $\langle\chi^2\rangle$ in such case boils down to integrating out the $\chi_k$ modes with momentum below the cutoff $k=He^{Ht}$. Since we have introduced the coarse-graining scale $k_s$ for the massless field $\phi$, we need to compute $\langle\chi^2\rangle$ on that scale in order to substitute in Eq.~(\ref{eq17}). Therefore, we multiply by $\theta(k_s-k)$ in the momentum integral to filter the appropriate superhorizon modes. Further neglecting the contribution to $\langle\chi^2\rangle$ from the pre-inflationary epoch ($\langle\chi^2(0)\rangle$) we have
\[\label{eq35}
\langle\chi^2\rangle_{\rm IR}=\int_{k=H}^{k=He^{Ht}}\frac{d^3\bk}{(2\pi)^3}\,\theta(k_s-k)|\chi_k(t)|^2\,.
\]
Using Eq.~(\ref{eq38}) we obtain
\[\label{eq12}
\langle\chi^2\rangle_{\rm IR}\simeq\theta(t-t_d)\left(\frac{H}{2\pi}\right)^2\frac{1+\coth[|\nu|\pi]}{2|\nu|}\frac{\epsilon^3}3
\left(1-\frac{1}{\epsilon^3a^3}\right)\,,
\]
where $t_d=H^{-1}\log \epsilon^{-1}$ is the time necessary for horizon-size modes to exceed the coarse-graining scale. Since $t_d$ is typically a few Hubble times, in the following we consider that $t\gg t_d$ at all relevant times. Also, using $|\nu|\simeq m_\chi/H$, one finds that Eq.~(\ref{eq12}) simplifies to
\[\label{eq5}
\langle\chi^2(t)\rangle_{\rm IR}\simeq\frac13\left(\frac{H}{2\pi}\right)^2\frac{H}{m_\chi}\,\epsilon^3\,.
\]
An equivalent manner to get to this result is by using the perturbation spectrum of a heavy field \cite{Riotto:2002yw}
\[\label{eq37}
{\cal P}_\chi(k)=\left(\frac{H}{2\pi}\right)^2\frac{H}{m_\chi}\left(\frac{k}{aH}\right)^3
\]
to compute $\langle\chi^2\rangle_{\rm IR}=\int P_\chi(k)\,d\ln\,k$. Owing to the scale-dependence, this equivalent approach leaves manifest that the main contribution to $\langle\chi^2\rangle_{\rm IR}$ stems from modes with wavelengths comparable to the coarse-graining scale, whereas the contribution from modes with larger wavelengths is suppressed. It is then clear that the scale-dependence featured in Eq.~(\ref{eq37}) is the reason of the factor $\epsilon^3$ in Eq.~(\ref{eq5}).

From Eq.~(\ref{eq5}) it follows that $m_\phi^2=g^2\langle\chi^2\rangle_{\rm IR}\ll H^2$, thus allowing us to neglect $\ddot \phi$ in Eq.~(\ref{eq17}). Also, in the limit $\bar m_\chi\ll g\phi$ we have $\langle\chi^2\rangle_{\rm IR}\propto |\phi|^{-1}$, hence quantum fluctuations of the $\chi$ field give rise to a slightly tilted linear effective potential for $\phi$. Quantum fluctuations of the $\Phi$ field also couple to $\chi$, which could result in dissipation effects further altering the dynamics of $\phi$. We discuss this possibility in the next section.

\subsection{On dissipation effects}
To discuss dissipation effects we bear in mind that the fields $\Phi$ and $\chi$ are to be eventually identified with fields in a supersymmetric theory, in the context of which dissipation effects have been extensively studied (see Ref.~\cite{Berera:2008ar} and references therein). In the particular case of MSSM flat directions, dissipation effects were studied in Ref.~\cite{Kamada:2009hy} and their importance have been recently emphasized in Ref.~\cite{Kawasaki:2012bk}. In this section we thus regard the system in Eq.~(\ref{eq16}) as contained in the scalar sector of a more general supersymmetric Lagrangian supporting the dissipation mechanisms studied in the literature. In this sense, a dissipation two-staged mechanism commonly considered in the literature couples the inflaton field to a intermediate heavy scalar field $\chi$. In turn, the heavy scalar $\chi$ decays into light degrees of freedom at a rate larger than the expansion rate. To give an example, in the context of supersymmetry such an interaction structure can be accounted for by the superpotential \cite{Berera:2008ar}
\[
W=g\Phi X^2+hXY^2~,
\]
where $\Phi$, $X$ and $Y$ are superfields and $\phi$, $\chi$, and $y$ refer
to their bosonic components. During inflation, the field y and its fermionic partner $\bar y$
remain massless, whereas the field $\chi$ and its fermion
partner $\psi_\chi$ obtain their masses through their couplings to
$\phi$, namely $m_{\psi_\chi}=m_\chi=g\phi$. In the low-temperature regime, defined by $m_\chi>T$, the dissipation coefficient is of the form
\[
\Upsilon\simeq C_\phi T^3/\phi^2\,,
\]
where $C_\phi\sim h^4{\cal N}_\chi{\cal N}_{\rm decay}^2$, ${\cal N}_\chi$ is the multiplicity of the $X$ superfield and ${\cal N}_{\rm decay}$ is the number of decay channels available in $X's$ decay. Although in our case the field $\phi$ does not play the role of the inflaton, the dissipation mechanism just described equally applies to $\phi$ since in the regime of interest to us $\chi$ is a heavy field.

Dissipation effects alter the dynamics of the coarse-grained field in the strong dissipation regime, defined by the condition $\Upsilon>3H$. Using the slow-roll equations of motion for $\phi$ and the radiation density \cite{Berera:2008ar} and the effective potential $V_{\rm eff}(\phi)=\frac12\,g^2\langle\chi^2\rangle_{\rm IR}\phi^2$, it can be shown that the condition $\Upsilon>3H$ translates into the bound $C_\phi\gg (g\epsilon^3)^{-3/2}\left(\phi/H\right)^2$. Taking typical values of $g$ and $\epsilon$ and using $\phi>H$, the resulting value for $C_\phi$ is very large indeed. We thus conclude that dissipation effects can be safely neglected unless the $\chi$ field belongs to a large representation of some Grand Unified Theory model \cite{BasteroGil:2006vr}. Owing to the coupling between $\phi$ and the radiation bath, dissipation effects can also affect the perturbation spectrum of $\phi$ even in the weak regime, defined by $\Upsilon\leq3H$. However, using again the slow-roll equations of motion for the field/radiation system it follows that the condition $\rho_r\gtrsim H^4$ (necessary for dissipative effects to alter the perturbation spectrum of $\phi$) translates into a bound on $C_\phi$ much tighter than the former one. In turn, this demands that the $\chi$ field belongs to a representation much larger than those considered in Ref.~\cite{BasteroGil:2006vr}. Therefore, we disregard dissipation effects on the perturbation spectrum of $\phi$ as well, which then allows us to use the correlation function in Eq.~(\ref{eq40}).

It is worth clarifying that dissipation effects are so suppressed in this case because their magnitude depends on the kinetic density of $\phi$, which in turn depends on the steepness of $V_{\rm eff}(\phi)$. Consequently, given the extreme flatness of the effective potential, dissipation effects can only become important when the smallness of the kinetic density of $\phi$ is counteracted by a large number of degrees of freedom taking part in the dissipation mechanism.

\subsection{Effective Fokker-Planck equation}\label{sub:FP}
After computing $\langle\chi^2\rangle_{\rm IR}$ we are ready to discuss the stochastic dynamics of long wavelength field $\phi$. Although Eq.~(\ref{eq17}) can be readily integrated to work out the two-point correlator $\langle\phi^2\rangle$, we find it more convenient for our purposes to write down and solve the Fokker-Planck equation (\ref{eq4}). Using $V'_{\rm eff}(\phi)=g^2\langle\chi^2\rangle_{\rm IR}\phi$ with $\langle\chi^2\rangle_{\rm IR}$ as given by Eq.~(\ref{eq5}) we have
\[\label{eq39}
\partial_tP=\frac{g^2H^2\epsilon^3}{36\pi^2}\,\partial_\phi\left(\frac{\phi P(\phi,t)}{(\bar m_\chi^2+g^2\phi^2)^{1/2}}\right)+
\frac{H^3}{8\pi^2}\,\partial_\phi^2P\,.
\]
At this point we emphasize that, contrary to the conventional expectation in stochastic inflation, the dynamics of the massless field $\phi$ depends on the auxiliary parameter $\epsilon\propto k_s$, thus acquiring a scale-dependence. However, such a behavior clearly originates from the coupling between $\phi$ and $\chi$, since the latter features a strong scale-dependence on superhorizon scales owing to its large mass (see Eqs.~(\ref{eq5}) and (\ref{eq37})). We wish to find an approximate solution to Eq.~(\ref{eq39}) for \mbox{$\bar m_\chi^2\ll g^2\phi^2$}. Restricting ourselves to $\phi>0$, the force provided by the effective potential is determined by
\[\label{eq20}
\frac{V'_{\rm eff}(\phi)}{3H}\simeq\frac{gH^2\epsilon^3}{36\pi^2}\,,
\]
which is constant. From this we may conclude that the motion of $\phi$ is always fluctuation dominated. This is because the magnitude of its classical motion per Hubble time is of the order $\Delta\phi\sim\dot\phi H^{-1}\sim V'_{\rm eff}/H^2\sim g\epsilon^3H\ll H$, and consequently much smaller than the typical amplitude of a field fluctuation during inflation. However, the relevant question to address is whether such a constant force driving the field to $\phi=0$ manages to restrain the field's diffusive motion. The answer to this question is straightforward: since the effective potential is a confining one in the entire domain of field values, the probability distribution becomes stationary after a sufficiently long wait \cite{Risken}. Although the approach to equilibrium in a linear potential can be studied analytically using the method of eigenfunctions \cite{Risken}, a simpler approach to the problem is possible because in the regime of interest to us ($0\leq \bar m_\chi\ll g\phi$) the Fokker-Planck equation can be integrated directly (equilibration timescales for the case of single field has recently been investigated in this context in \cite{Enqvist:2012xn}). Neglecting the term in $\bar m_\chi^2$ in Eq.~(\ref{eq39})  we have
\[\label{eq3}
\partial_tP=\frac{gH^2\epsilon^3}{36\pi^2}\,\partial_\phi P+\frac{H^3}{8\pi^2}\partial_\phi^2P\,.
\]
We emphasize that since this equation holds for $0\leq \bar m_\chi\ll g\phi$ only, we should not expect to recover the equilibrium distribution that follows from imposing $\partial_tP=0$ in Eq.~(\ref{eq39}) (valid for all $\phi$) by taking the limit $t\to\infty$ in the solution to Eq.~(\ref{eq3}). Nevertheless, as explained in the next section, the solution computed below is sufficient for our purposes.

To integrate Eq.~(\ref{eq3}) we impose a gaussian distribution with mean and variance\footnote{The usual initial condition $P(\phi,0)=\delta(\phi-\phi_0)$ is trivially recovered in the limit $\sigma_0=0$.}
\[\label{eq36}
\mu_0\equiv\langle\phi(0)\rangle=\phi_0\quad,\quad\sigma_0^2\equiv\langle(\phi(0)-\phi_0)^2\rangle
\]
as initial condition for $P(\phi,t)$. We then obtain
\[\label{eq14}
P(\phi,t)=\frac1{\sqrt{2\pi\sigma^2(t)}}\exp\left[-\frac{(\phi-\phi_0+\kappa t)^2}{2\sigma^2(t)}\right]\,,
\]
where
\[\label{eq25}
\sigma^2(t)\equiv\sigma_0^2+\frac{H^3t}{4\pi^2}\quad\textrm{and}\quad\kappa\equiv \frac{gH^2\epsilon^3}{36\pi^2}\,.
\]
From Eq.~(\ref{eq14}) it follows that the effect of the linear potential is a mere displacement of the center of the distribution. Consequently, the field remains gaussian distributed  with mean $\mu=\phi_0-\kappa t$ and variance $\sigma^2(t)$. Nevertheless, to understand the evolution of the distribution in terms of the model parameters it is convenient to introduce the timescales for the change of the variance $\sigma^2$, $\tau_\sigma$, and for the change of the mean field $\mu$, $\tau_\mu$. These are given by
\[\label{eq24}
\tau_\sigma\equiv\frac{\sigma^2}{\dot\sigma^2}=\left(1+\frac{4\pi^2\sigma_0^2}{H^2N}\right)H^{-1}N\simeq H^{-1}N\quad,\quad\tau_\mu\equiv\frac{\mu}{\dot\mu}=\left(\frac{36\pi^2}{g\epsilon^3}\frac{\phi_0}{H}\right)H^{-1}\gg H^{-1}\,.
\]
where $N\equiv Ht$ is the number of elapsed $e$-foldings. Also, we have neglected the initial variance $\sigma_0^2$ to estimate $\tau_{\sigma}$. The evolution of $P$ can be described now in terms of $\tau_\sigma$ and $\tau_\mu$.

When $\tau_\sigma\ll \tau_\mu$ the evolution of $P(\phi,t)$ is dominated by the diffusive motion, i.e. the variance $\sigma^2$ grows linearly with time (after neglecting $\sigma_0^2$) whereas the mean $\mu=\phi_0-\kappa t$ remains virtually unchanged since $\tau_\sigma\ll\tau_\mu$ can be recast as $\kappa t\ll\phi_0$. This implies that  the coupling between $\phi$ and $\chi$ is unable to restrain the random motion of the former on timescales of the order $\tau_\sigma$. As a result, the fluctuations of the field correspond to a massless free field. In terms of elapsed $e$-foldings, the diffusive evolution of $P(\phi,t)$ can be considered a good approximation to its dynamics provided $N\ll N_{\rm drift}$, where
\[\label{eq30}
N_{\rm drift}\equiv\frac{36\pi^2}{g\epsilon^3}\frac{\phi_0}{H}=\frac{H^2}{\kappa}\,\frac{\phi_0}{H}\,.
\]
Using typical values of the parameters $g\lesssim1$, $\epsilon\ll 1$ and assuming $\phi_0>H$, we conclude that such epoch encompasses an extremely large phase of inflation. However, since $\tau_\sigma$ increases with time while $\tau_\mu$ remains constant, the diffusive motion of $P$ does not persist indefinitely. If inflation is sufficiently long-lasting so that $N\gg N_{\rm drift}$, then $\tau_\mu\ll\tau_\sigma$. In that case, the distribution drifts towards $\phi=0$ on the timescale $\tau_\mu$ whereas its variance remains approximately constant. Consequently, and despite the fluctuation dominated evolution on the Hubble timescale, the motion of the field becomes potential dominated on the timescale $\tau_\mu\gg H^{-1}$. Although the length of inflation required for that to happen is extremely large indeed, this could be the case provided eternal inflation is considered \cite{Linde:1986fc}.

The above results indicate that even in the presence of an effective mass $m_\phi^2=g^2\langle\chi^2\rangle_{\rm IR}$ the field $\phi$ manages to fluctuate as a massless, free field. The fundamental feature that allows $\phi$ to fluctuate freely is that $m_\phi^2$ is inverse to $\phi$ [c.f. Eq.~(\ref{eq5})]. The key to understand this result physically relies in the heaviness of the $\chi$  field. As long as $m_\chi^2\simeq g^2\phi^2\gg H^2$, the response of the system to a random fluctuation increasing $\langle\phi^2\rangle$ is to adjust $m_\phi^2=g^2\langle\chi^2\rangle_{\rm IR}$ to a smaller value according to Eq.~(\ref{eq5}) in much less than a Hubble time. This is to say that the swift compliance of the system to adjust $\langle\chi^2\rangle_{\rm IR}$ allows the mean square $\langle\phi^2\rangle$ to grow unimpeded.

\subsection{Below the crossover scale}\label{sec.crossover}
The regime of free fluctuations that we have just described cannot account for the evolution of $\phi$ indefinitely. Owing to its random motion, the expectation value of $\phi$ can grow large in some coarse-graining patches, thus making $\chi$ heavier and keeping our computation valid. However, in some patches $\phi$ takes on progressively smaller values. Eventually, the field $\phi$ decreases enough so that $\chi$ ceases to be a heavy field, which then gives rise to the production of superhorizon fluctuations of $\chi$. In turn, this production enhances the effective potential of $\phi$ making it steeper. The field $\phi$ is consequently driven closer to the origin where $\chi$ is even lighter. As a result of this, the production of superhorizon fluctuations reinforces and accumulates as the spectrum of the $\chi$ field becomes less scale-dependent. Since this implies that the $\phi$ is pushed harder towards $\phi=0$, the final outcome of the production process is that the motion of $\phi$ becomes confined around the origin, thus terminating the phase of free fluctuations.

This trapping of the field $\phi$ occurs similarly to the rapid decay of the inflaton field at the end of the broad parametric resonance regime \cite{Kofman:1994rk}, or similarly to the moduli trapping mechanism \cite{Kofman:2004yc} (see also \cite{Green:2009ds,Barnaby:2009mc}). Nevertheless, in this paper we do not concern ourselves with the details of the trapping phase, but rather focus on the ramifications of the existence of a scale below which the motion of $\phi$ becomes confined around $\phi=0$. In the following we refer to this as the crossover scale and denote it by $\phi_c$. We further assume that once the field falls below the crossover scale and becomes trapped at the origin, no random fluctuation of the system can take the field back to $\phi>\phi_c$.
\begin{figure}[t]
\centering\epsfig{file=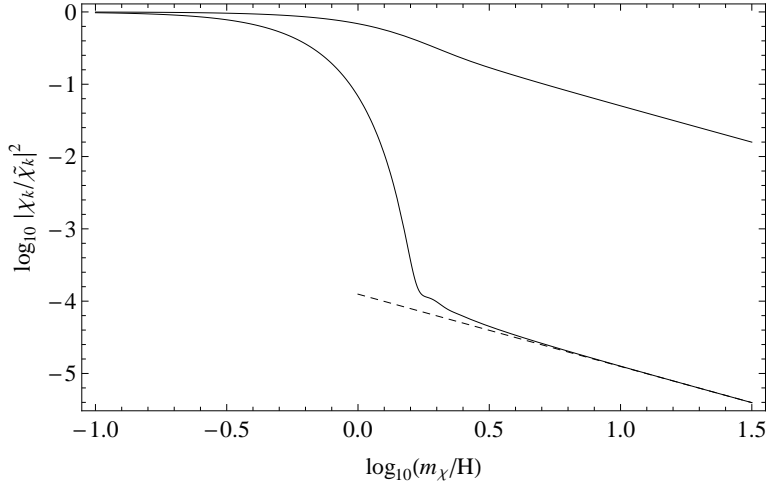,width=10.0cm}\caption{Plot of $|\chi_k/\widetilde\chi_k|^2$ vs. $m_\chi$. Solid curves from top to bottom correspond to the time when the $k$-mode exits the horizon and when the mode exits the coarse-graining scale. The dashed line represents the plotted ratio in the limit $m_\chi\gg H$, i.e. $(H/m_\chi)\,\epsilon^3$. The plot is built using $\epsilon=5\times10^{-2}$.}\label{fig1}
\end{figure}

From the previous discussion it is clear that the crossover scale is necessarily related to the Hubble scale, for $\chi$ must be a heavy field for $\phi$ to fluctuate quasi-freely. Therefore, it seems reasonable to anticipate that setting $\phi_c$ large enough so that $g\phi_c$ is a few times larger than $H$ should suffice to guarantee that $\phi$ performs quasi-free fluctuations when $\phi\geq\phi_c$. However, it is still possible to have a more precise estimate of $\phi_c$. As explained before, the crucial feature leading to the diffusive motion of $\phi$ is the inverse dependence $\langle\chi^2\rangle_{\rm IR}\propto \phi^{-1}$, for in that case its effective potential does not become steeper as $\phi$ grows large. This observation constitutes the basis for our estimate, which we explain below.

We begin by setting $\bar m_\chi=0$ so that $m_\chi=g\phi$. To identify the smallest scale at which $\langle\chi^2\rangle_{\rm IR}$ becomes approximately inverse to $\phi$ we utilize the scale-dependence of the perturbation spectrum of a heavy field (see Eq.~(\ref{eq37})). Such a scale-dependence implies that the quantum average $\langle\chi^2\rangle_{\rm IR}$ on the coarse-graining scale is mainly determined by the amplitude of the modes $\chi_k$ on that scale. This can be clearly appreciated from Eq.~(\ref{eq5}) which, apart from constant factors, can be understood as the product of $|\chi_k|^2$,  as given by Eq.~(\ref{eq38}), times $k_s^3$. Using this, we estimate the crossover scale $\phi_c$ as the scale at which the amplitude $|\chi_k|^2$, rather than $\langle\chi^2\rangle_{\rm IR}$ itself, becomes approximately inverse to $m_\chi=g\phi$. In Fig.~\ref{fig1} we depict the ratio $|\chi_k/\widetilde\chi_k|^2$, where $\widetilde\chi_k$ is the solution to Eq.~(\ref{eq32}) for an exactly massless field, for a given $k$-mode as a function of $m_\chi$. We plot this ratio at the time of horizon exit (upper, solid curve), and when the $k$-mode exits the coarse-graining scale (lower, solid curve). The dashed curve corresponds to the limit of the plotted ratio at very large mass, namely
\[
\left|\chi_k/\widetilde \chi_k\right|^2\to\frac{H}{m_\chi}\,\epsilon^3\quad\textrm{for}\quad m_\chi\gg H\,,
\]
where we evaluate the ratio at the coarse-graining scale $k_s$. We define the crossover scale $\phi_c$ as the smallest value of $\phi$ at which the ratio $|\chi_k/\widetilde \chi_k|^2$ approximately matches (up to the first derivative) its value in the limit $m_\chi\gg H$. Eye inspection of Fig.~\ref{fig1} suffices to estimate this scale just by reading $m_\chi$ when the bottom, solid curve and its slope match those of the dashed one. This roughly gives
\[\label{eq29}
\phi_c\simeq\sqrt{10}\,H/g\,.
\]
If there is a non-negligible contribution $\bar m_\chi\neq 0$, the crossover scale must be set sufficiently high so that $\bar m_\chi\ll g\phi_c$ and $\langle\chi^2\rangle_{\rm IR}$ becomes approximately inverse to $\phi$ for $\phi\geq\phi_c$, thus allowing the virtually unimpeded random motion of $\phi$.

We wish to remark that the above is not intended to be an accurate computation, but a mere guess based on what is required for the coupled massless field $\phi$ to fluctuate freely.

\section{General expression for field fluctuations}\label{sec.fluct}
Having elucidated the growth of fluctuations in the large field regime ($\phi\gg H/g$), we proceed now to describe the probability density in the range $\phi\geq\phi_c$. As discussed before, if the initial distribution peaks around $\phi=\phi_0$ well above the crossover scale, the field manages to perform a quasi-free random motion (for $N\ll N_{\rm drift}$). Owing to this, the expectation value of $\phi$ can grow small and approach the crossover scale in some coarse-graining patches.

As argued in Sec.~\ref{sec.crossover}, whenever the field's expectation value falls below $\phi_c$ its evolution becomes confined around $\phi=0$ indefinitely. It then follows that the probability density describing the random motion of $\phi$ for $\phi\geq\phi_c$ is not conserved with time, and hence it cannot be properly described by Eq.~(\ref{eq14}). To find the appropriate probability density function we invoke our assumption on the indefinite trapping of the field around $\phi=0$. Such an assumption allows us to understand the stochastic dynamics of $\phi$ as that of a brownian particle performing a drift-diffusion motion for $\phi>\phi_c$ in the presence of an absorbing barrier located at $\phi=\phi_c$. This implies that the probability density is determined by the equation
\[\label{eq26}
\partial_tP=\kappa\,\partial_\phi P+
\frac{H^3}{8\pi^2}\partial_\phi^2P\,,
\]
with the boundary condition accounting for an absorbing barrier at $\phi=\phi_c$. Such a condition is of the form \cite{Chandrasekhar:1943ws,Cox,Molini}
\[\label{eq21}
P(\phi_c,t)=0\,,
\]
and has been previously used in the context of stochastic inflation in Ref.~\cite{Lorenz:2010vf}. Imposing a gaussian distribution with mean $\phi_0$ and variance $\sigma_0^2$ as initial condition, the solution to Eq.~(\ref{eq26}) for $\phi>\phi_c$ (where $P$ is positive) is found to be
\[\label{eq6}
P(\phi,t)=\frac{1}{\sqrt{2\pi\sigma^2}}\left\{\exp\left[-\frac{(\phi-\varphi_+)^2}{2\sigma^2}\right]
-C\exp\left[-\frac{(\phi-\varphi_-)^2}{2\sigma^2}\right]\right\}\,,
\]
where
\[\label{eq28}
\varphi_+\equiv\phi_0-\kappa t\quad,\quad\varphi_-\equiv-\phi_0+2\phi_c-\kappa t-\frac{8\pi^2\sigma_0^2\kappa}{H^3}
\]
and $C$ is a constant given by
\[
C\equiv\exp\left[\displaystyle\frac{8\pi^2\kappa}{H^3}\left(\phi_0-\phi_c+
\frac{4\pi^2\sigma_0^2\kappa}{H^3}\right)\right]\,.
\]
The first term in Eq.~(\ref{eq6}) corresponds to the probability density in Eq.~(\ref{eq14}), whereas the second term accounts for the loss of probability due to the absorbing barrier at $\phi=\phi_c$.

\begin{figure}[htbp]
\centering\epsfig{file=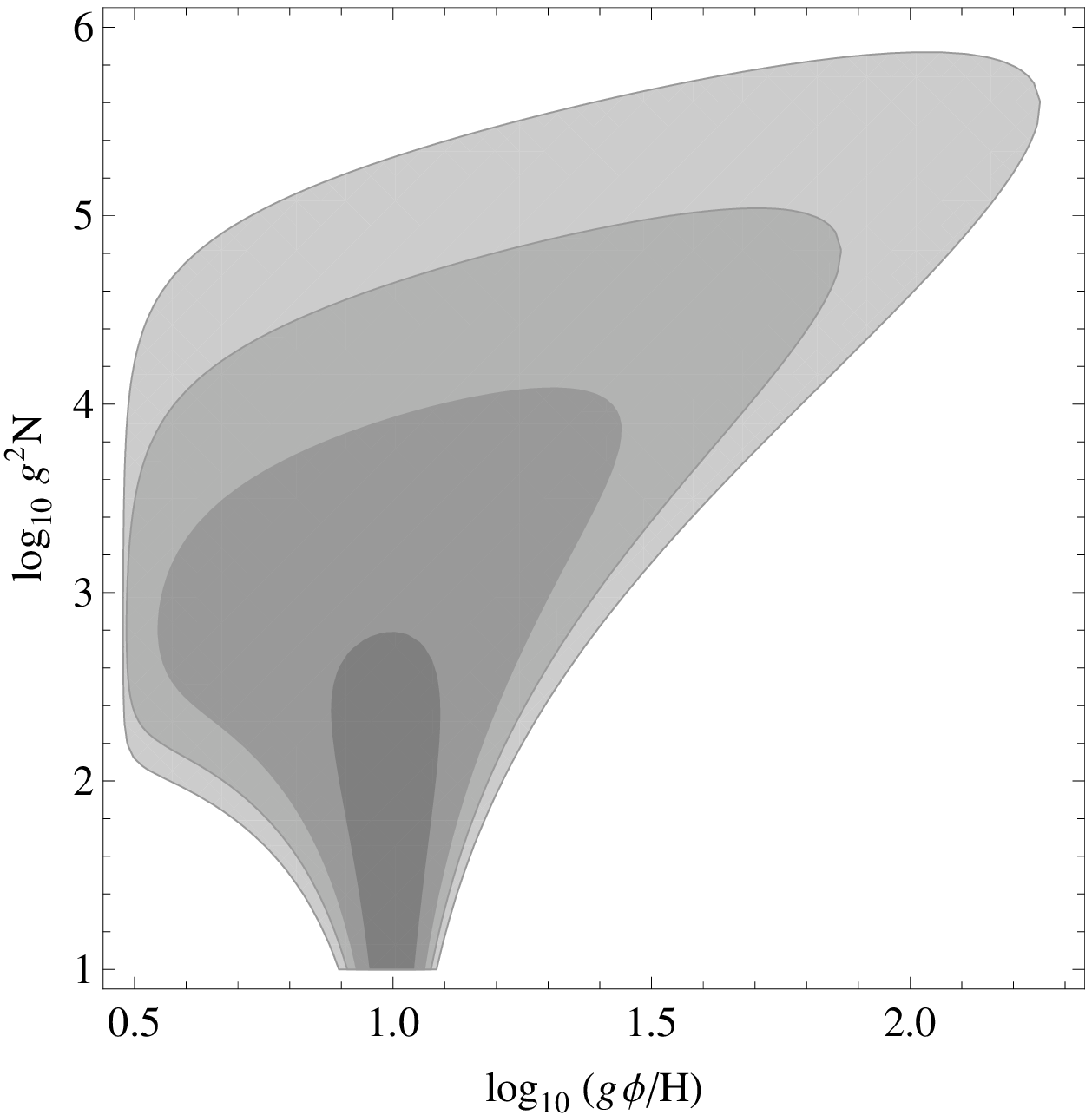,width=7.5cm}\hspace{0.25cm}\epsfig{file=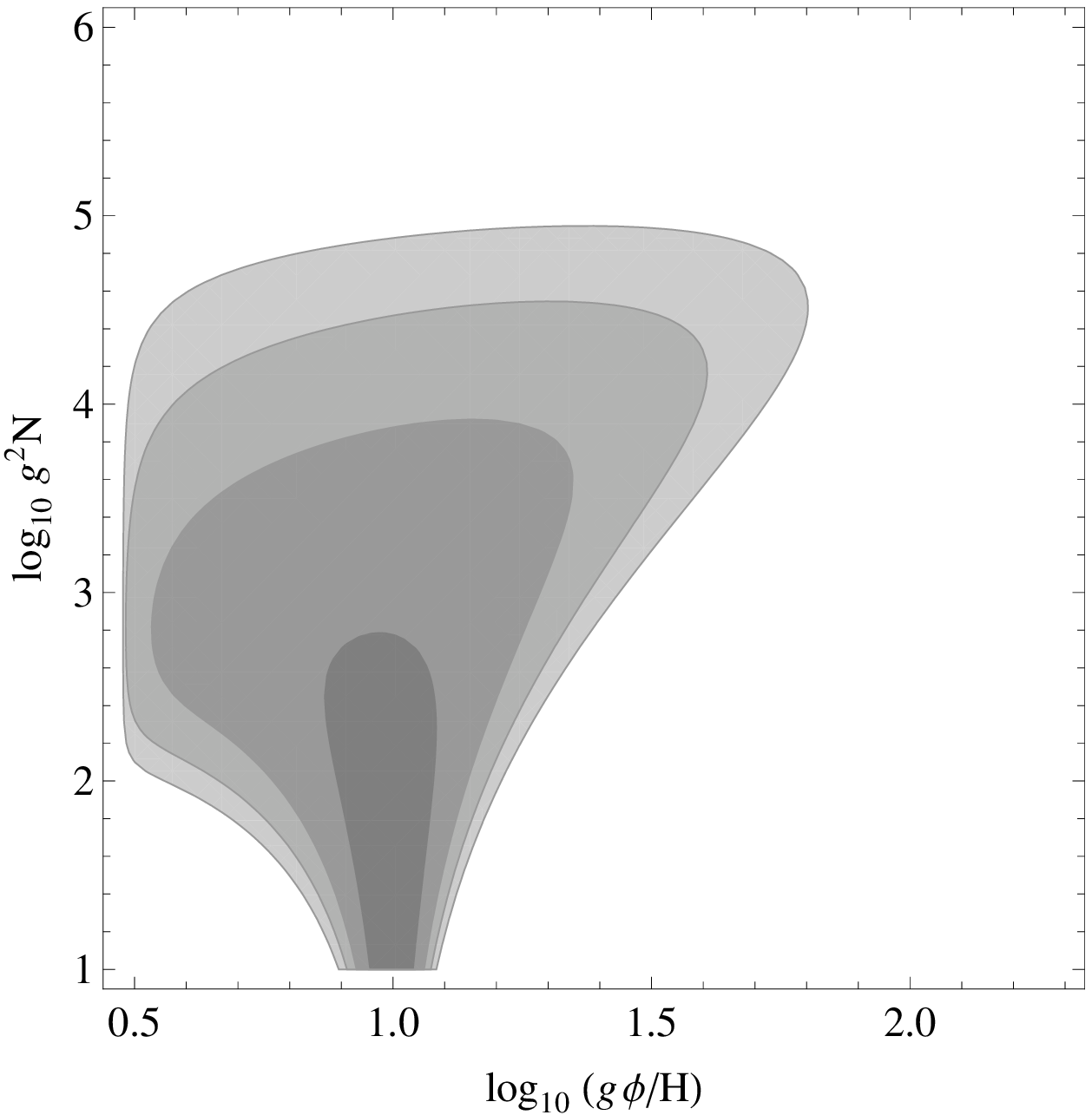,width=7.5cm}
\caption{Contours of $g^{-1}P(\phi,t)$ (see Eq.~(\ref{eq6})). The shaded areas correspond to $g^{-1}P(\phi,t)= 10^{-\gamma}$ for $\gamma\leq1,2,3,4$ (from heavier to lighter shading). To illustrate the behavior of $P(\phi,t)$ we have chosen $\phi_0=10H/g$, $\phi_c=3H/g$ and $\sigma_0=5\times10^{-2}H/g$. To demonstrate the effect of the parameter $\kappa$ we choose $\kappa=10^{-4}gH^2$ in the left-hand panel and $\kappa=10^{-3}gH^2$ in the righthand one.}\label{fig4}
\end{figure}

The behavior of $g^{-1}P(\phi,t)$ is illustrated in Fig.~\ref{fig4}, where we plot contours enclosing the region $g^{-1}P(\phi,t)=10^{-\gamma}$ for $\gamma\leq1,2,3,4$ (from heavier to lighter shading) using $\phi_0=10H/g$, $\phi_c=3H/g$ and $\sigma_0=5\times10^{-2}H/g$. In the left-hand panel we take $\kappa=10^{-4}gH^2$ and $\kappa=10^{-3}gH^2$ in the righthand one to appreciate the effect of the potential tilt $\kappa$ on the probability density. The plot shows that for larger values of $\kappa$, implying a steeper linear potential, the field distribution is ``pushed'' towards the absorbing barrier at an earlier time.

\subsection{Fraction of survivors}
\begin{figure}[htbp]
\centering\epsfig{file=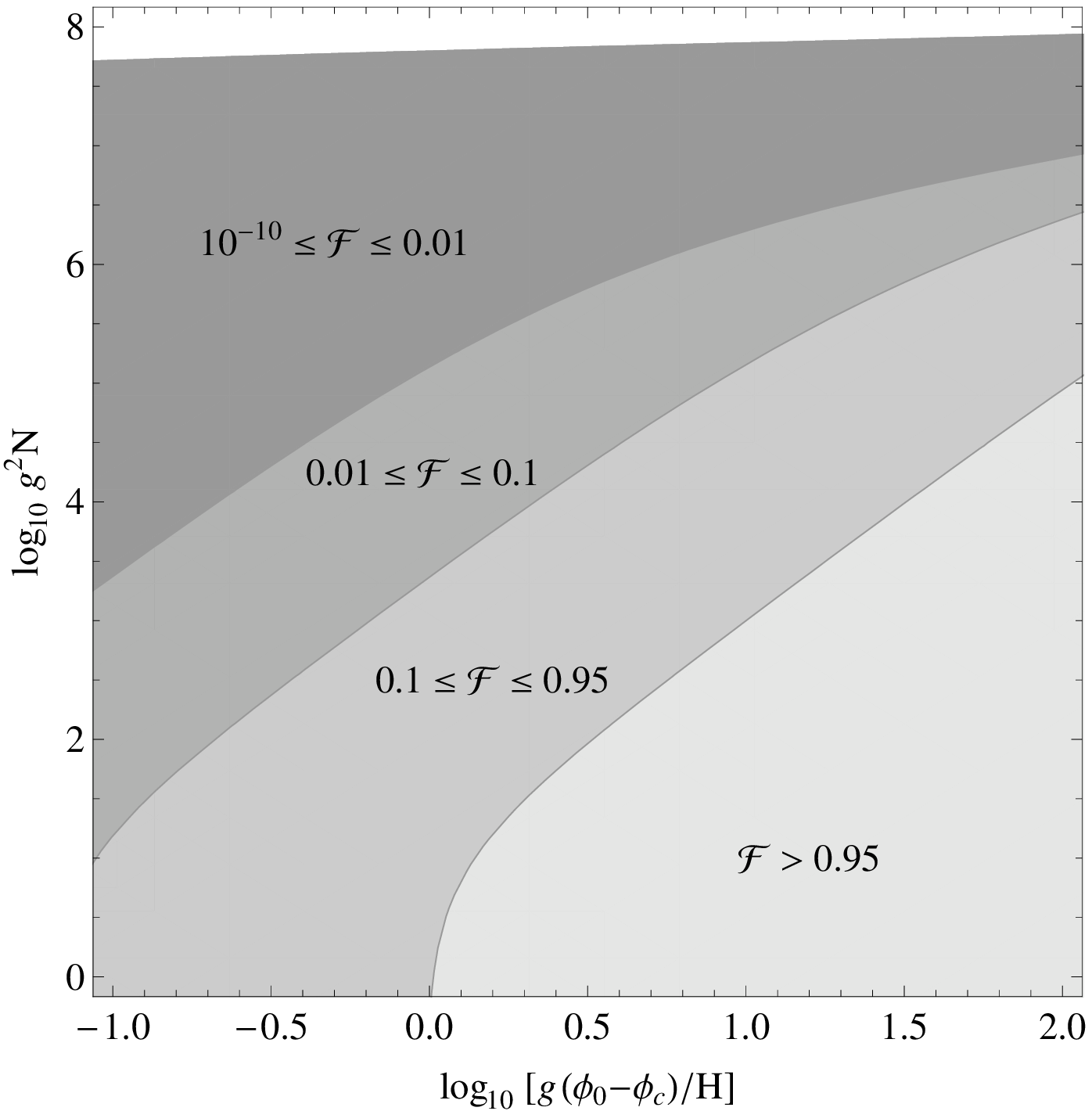,width=7.5cm}\hspace{0.25cm}\epsfig{file=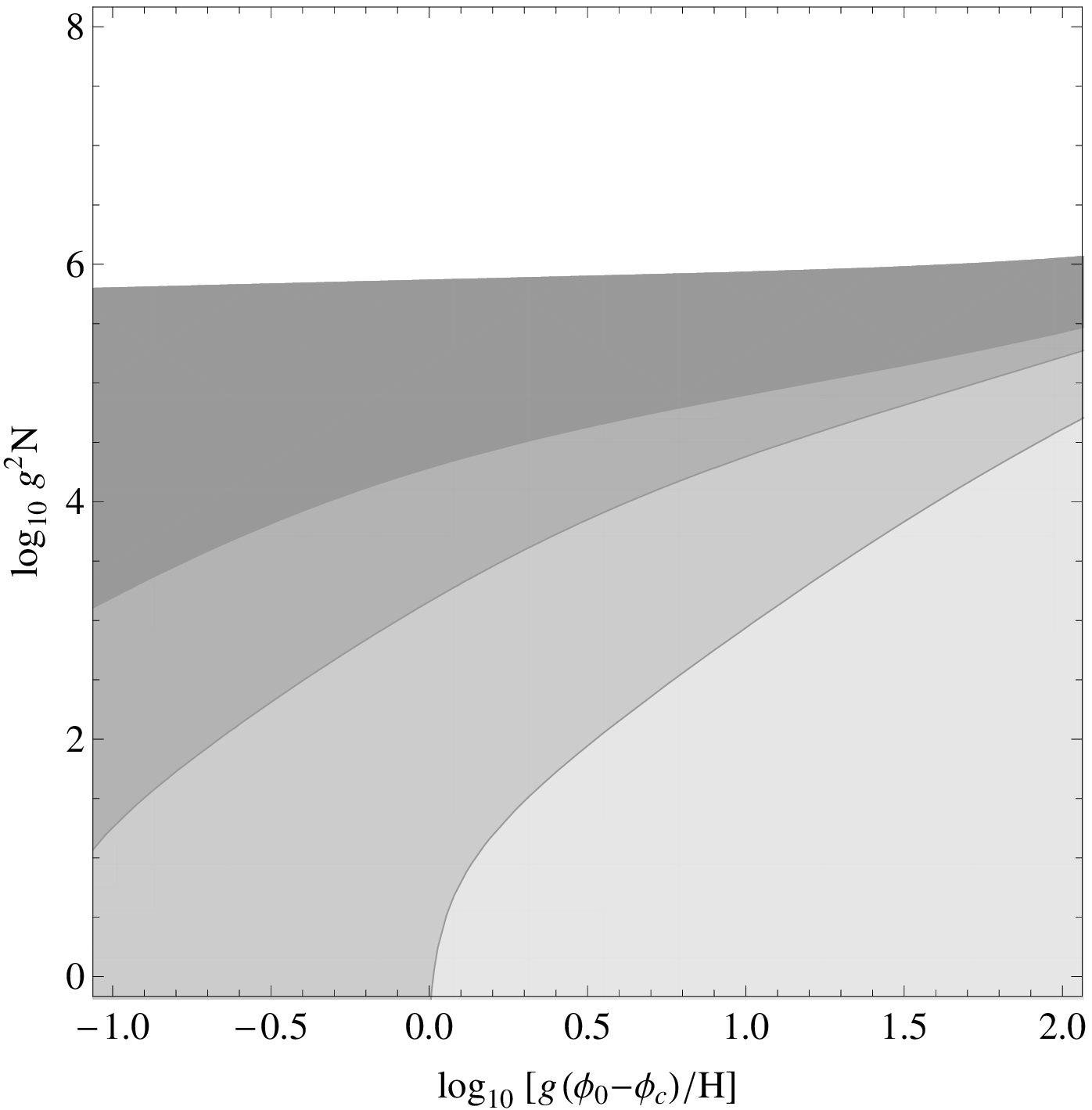,width=7.5cm}
\caption{Contours of the fraction ${\cal F}$ varying within certain range (as indicated in the plot). We use \mbox{$\sigma_0=0.5H/g$} and $\kappa=10^{-4}gH^2$ for the left-hand panel and $\kappa=10^{-3}gH^2$ in the righthand panel, where we plot the same contours, to demonstrate the effect of the tilt $\kappa$.}\label{fig5}
\end{figure}

In order to trace the approach to the equilibrium distribution we consider the fraction of the ensemble in which the field is above the crossover scale, i.e. fluctuating with an out-of-equilibrium amplitude, after $N$ $e$-foldings of inflation. This fraction, which we denote by ${\cal F}$, can be readily computed by integrating Eq.~(\ref{eq6}):
\[\label{eq33}
{\cal F}(t)=\int_{\phi_c}^\infty P(\phi,t)\,d\phi=\frac{1}{\sqrt{2\pi\sigma^2}} \left[I_1(\varphi_+,t)-CI_1(\varphi_-,t)\right]\,,
\]
where
\[
I_1(\varphi)=\int_{\phi_c}^{\infty}\exp\left[-\frac{(\phi-\varphi)^2}{2\sigma^2}\right]d\phi=
\sqrt{\frac{\pi\sigma^2}2}\left[1+{\rm Erf}\left(\frac{\varphi-\phi_c}{\sqrt{2\sigma^2}}\right)\right]\,.
\]
Of course, the fraction of the ensemble in which the field fluctuates with the equilibrium amplitude after $N$ $e$-foldings is $1-{\cal F}(t)$. If inflation lasts for a number of $e$-foldings $N\ll N_{\rm drift}$ we can neglect the drift term and consider $\kappa\simeq0$. In that case, Eq.~(\ref{eq33}) simplifies to
\[
{\cal F}(t,\kappa=0)={\rm Erf}\left(\frac{\phi_0-\phi_c}{\sqrt{2\sigma^2}}\right)\,.
\]
In Fig.~\ref{fig5} we represent several regions in which ${\cal F}$ varies within certain range (as indicated). Similarly to Fig.~\ref{fig4}, we use $\kappa=10^{-4}gH^2$ in the left-hand panel and $\kappa=10^{-3}gH^2$ in the righthand one. Again, the plot shows that as the tilt $\kappa$ increases the field distribution is pushed towards the equilibrium state at an earlier time.

\subsection{Out-of-equilibrium fluctuations}
Next, we compute the variance $\langle\phi^2\rangle$, which we write as
\[\label{eq31}
\langle\phi^2\rangle\equiv\langle\phi^2\rangle_{\rm eq}+\langle\phi^2\rangle_{\rm fl}\,,
\]
where $\langle\phi^2\rangle_{\rm eq}$ is the  contribution from equilibrium fluctuations, which the field is assumed to attain after falling below the crossover scale, and $\langle\phi^2\rangle_{\rm fl}$ is the contribution from the fraction of the ensemble in which $\phi$ undergoes quasi-free fluctuations. Using Eq.~(\ref{eq6}), the latter can be readily computed as
\[\label{eq27}
\langle\phi^2\rangle_{\rm fl}=\int_{\phi_c}^\infty \phi^2P(\phi,t)\,d\phi=\frac1{\sqrt{2\pi\sigma^2}}\left[I_2(\varphi_+)-CI_2(\varphi_-)\right]\,,
\]
where
\begin{eqnarray}
I_2(\varphi)&=&\int_{\phi_c}^\infty\phi^2\exp\left[-\frac{(\phi-\varphi)^2}{2\sigma^2}\right]\,d\phi
\nonumber\\
&=&\exp\left[-\frac{(\varphi-\phi_c)^2}{2\sigma^2}\right]\sigma^2(\varphi+\phi_c)
+\sqrt{\frac{\pi\sigma^2}{2}}\left(\sigma^2+\varphi^2\right)\left[1+{\rm Erf}\left(\frac{\varphi-\phi_c}{\sqrt{2\sigma^2}}\right)\right].
\end{eqnarray}

To work with a simple, manageable expression we consider an inflationary phase of length $N\ll N_{\rm drift}$, in which case we can take $\kappa=0$. As previously noticed, this is the most plausible case unless eternal inflation, or some other long-lasting inflationary scenario, is considered. Using Eq.~(\ref{eq28}) with $\kappa=0$ and Eq.~(\ref{eq27}) we find
\begin{equation}\label{eq8}
\langle\phi^2\rangle_{\rm fl}=\exp\left(-\frac{\delta^2}{2\sigma^2}\right)\sqrt{\frac{2}{\pi}}\,\sigma\delta+2\phi_c\delta
+\left(\sigma^2+\delta^2+\phi_c^2\right){\rm Erf}\left[\frac{\delta}{\sqrt{2\sigma^2}}\right]\,,
\end{equation}
where we introduce the initial departure from the crossover point as $\delta\equiv\phi_0-\phi_c$ for convenience.

If $\phi$ begins above the crossover scale in the entire ensemble ($\phi_0\gg\phi_c$ and  $\sigma_0^2\ll\delta^2$), then $\langle\phi^2\rangle_{\rm fl}$ should grow linearly with time. Our Eq.~(\ref{eq8}) must then reproduce the epoch of linear growth with time for as long as $\sigma^2\ll\delta^2$. Using that ${\rm Erf}(x\gg1)\approx1$ and that $\langle\phi^2(0)\rangle=\sigma_0^2+\phi_0^2$ we find
\[\label{eq9}
\langle\phi^2(t)\rangle_{\rm fl}\simeq2\phi_c\delta+\left(\sigma^2+\delta^2+\phi_c^2\right)
=\langle\phi^2(0)\rangle+\frac{H^3t}{4\pi^2}\,,
\]
which is the expected result \cite{Linde:1982uu,Starobinsky:1982ee,Vilenkin:1982wt}. Alternatively, this result can be derived from Eq.~(\ref{eq8}) simply by taking the absorbing barrier to infinity, i.e. $\phi_c\to-\infty$, or $\delta\to\infty$ equivalently. Next, we obtain the behavior of $\langle\phi^2\rangle_{\rm fl}$ at later times, when the field becomes trapped below $\phi_c$ in a substantial fraction of the ensemble. Such a situation corresponds to $\sigma^2\gg\delta^2$. Since $g\phi_0> H$ and $g\phi_c\sim H$ (see Eq.~(\ref{eq29})), the condition $\sigma^2\gg\delta^2$ implies both $\sigma^2\gg\phi_c^2$ and $\sigma^2\gg \sigma_0^2$. Further expanding ${\rm Erf}(|x|\ll1)\simeq2\pi^{-1/2}x$ in Eq.~(\ref{eq8}) we obtain
\[\label{eq34}
\langle\phi^2(t)\rangle_{\rm fl}\simeq2
\delta\sqrt{\frac{H^3t}{2\pi^3}}\,.
\]
Comparing this expression with Eq.~(\ref{eq9}) we conclude that once $\phi$ falls below $\phi_c$ in a substantial part of the ensemble the growth of inhomogeneities proceeds at a rate slower than in the uncoupled case, with the scaling $\langle\phi^2\rangle_{\rm fl}\propto \sqrt{t}$. This slowdown is a direct consequence of the coupling between $\phi$ and $\chi$, which prevents the growth of the field fluctuations (away from the equilibrium value) below the crossover scale.
\begin{figure}[htbp]
\centering\epsfig{file=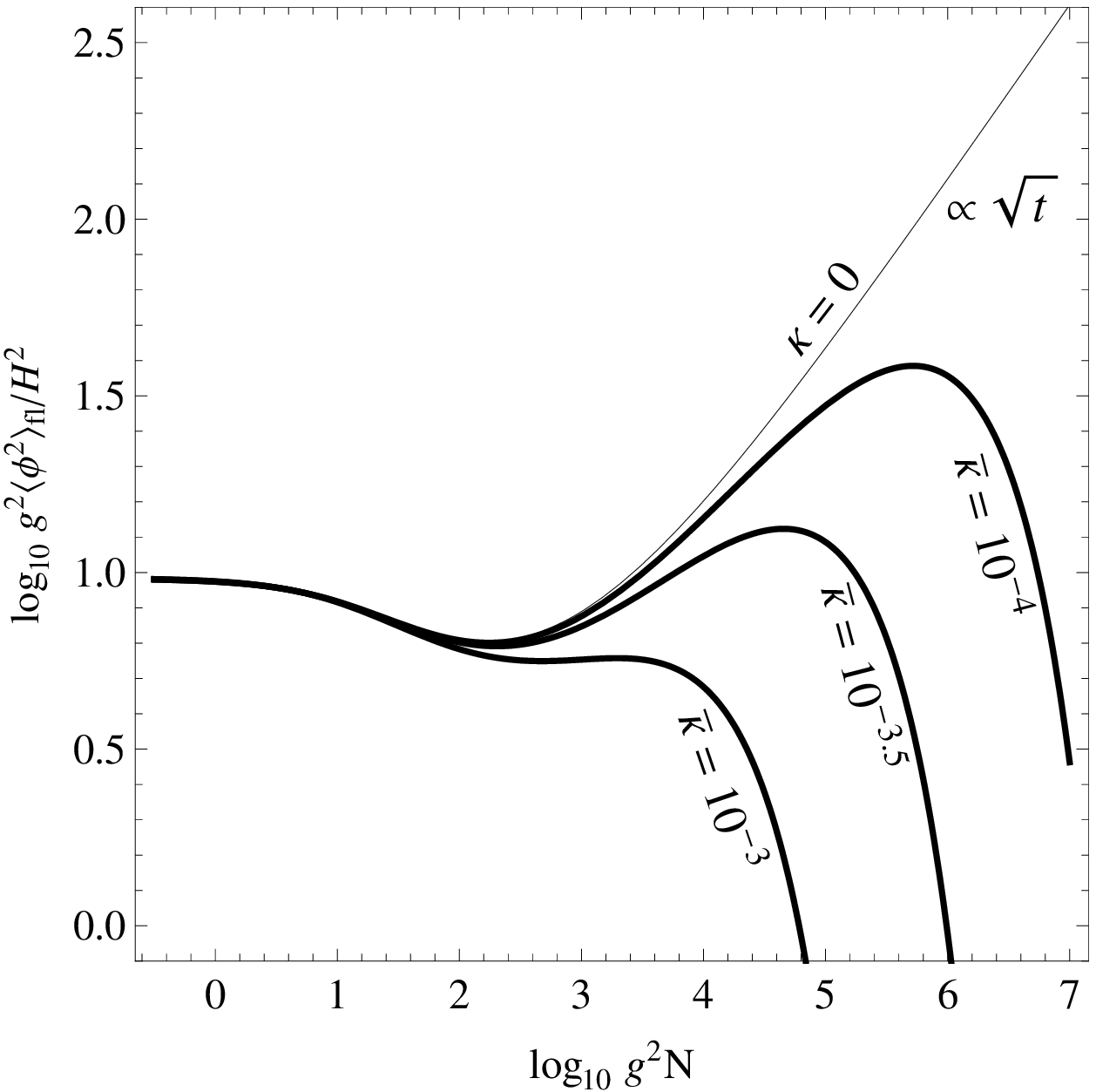,width=7.5cm}\hspace{0.25cm}\epsfig{file=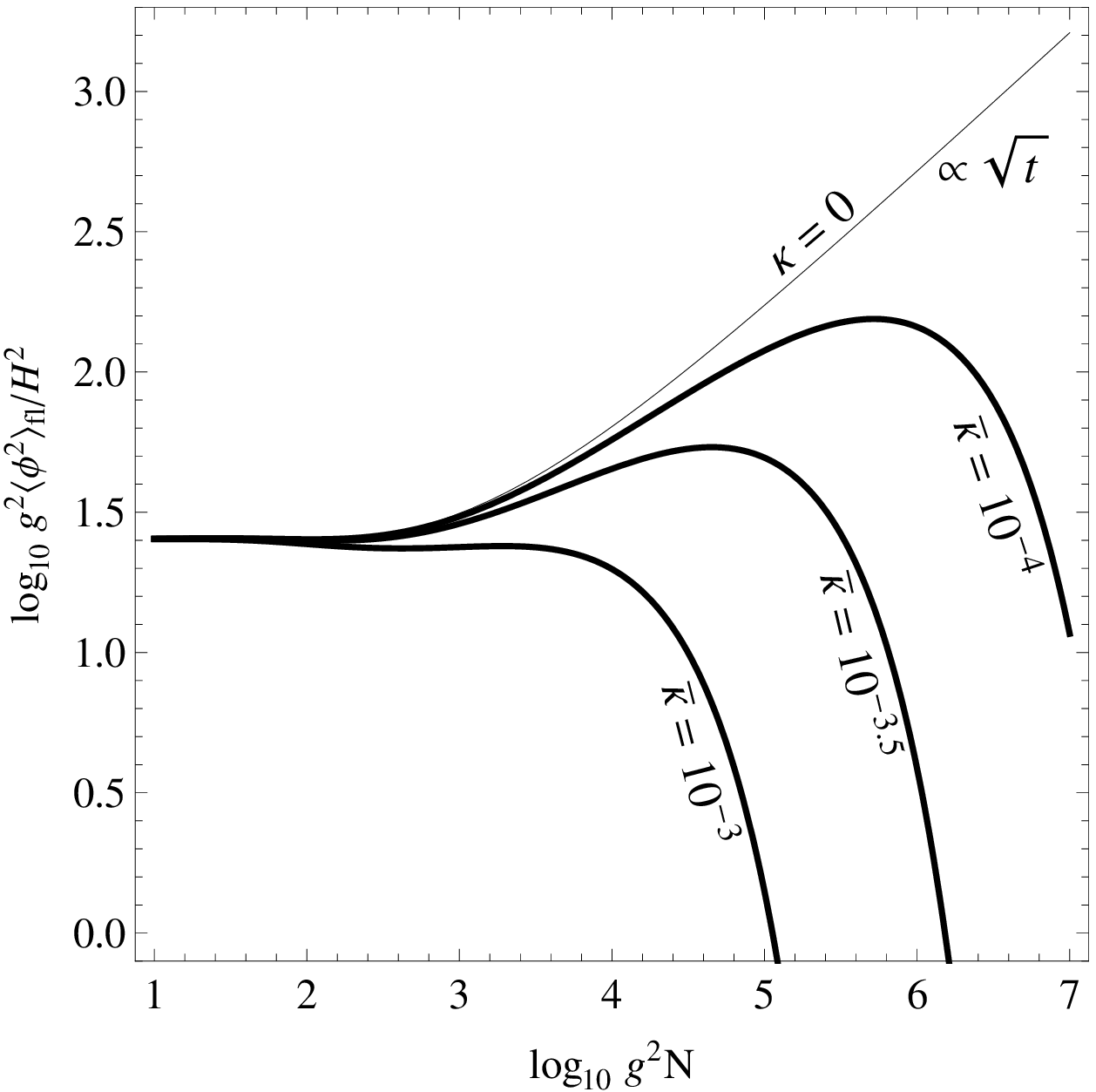,width=7.5cm}
\caption{Plot of $\langle\phi^2\rangle$ as a function of $g^2N$. We choose $\phi_c=3H/g$, $\sigma_0^2=0.25H^2/g^2$ and $\phi_0=3.5H/g$ for the left-hand panel and $\phi_0=5H$ for the righthand panel. The thin curve corresponds to $\kappa=0$ (Eq.~(\ref{eq8})) in which the evolution of the distribution is dominated by the diffusive motion. The thick curves correspond to the prediction in Eq.~(\ref{eq27}) for three different values of $\bar\kappa\equiv \kappa/(gH^2)$.}\label{fig6}
\end{figure}

The behavior of $\langle\phi^2\rangle_{\rm fl}$ for several values of the tilt parameter $\kappa$ is displayed in Fig.~\ref{fig6}. In the left-hand panel we show the case when the field begins slightly above the crossover scale. We take $\phi_c=3H/g$, $\phi_0=3.5H/g$ and $\sigma_0^2=0.25H^2/g^2$. For these values, a substantial fraction of the initial distribution begins with $\phi<\phi_c$. As a result, the probability of finding a freely fluctuating field in a member of the ensemble reduces, and so does $\langle\phi^2\rangle_{\rm fl}$ as a result. The thin curve represents the prediction in Eq.~(\ref{eq8}) corresponding to a negligible drift motion, and constitutes a good approximation to $\langle\phi^2\rangle_{\rm fl}$ as long as $N\ll N_{\rm drift}$. For $N\gg N_{\rm drift}$ the motion of the distribution is dominated by the drift, which then suppresses the out-of-equilibrium fluctuations. The thick curves in Fig.~\ref{fig6} correspond to the prediction in Eq.~(\ref{eq27}) for different values of $\kappa$ (as indicated in the plot). If inflation persists for long enough, the fraction of the ensemble with out-of-equilibrium field fluctuations becomes exponentially suppressed [c.f. Fig.~\ref{fig5}], hence the field fluctuations approach their equilibrium amplitude. In the righthand panel we change the initial value to $\phi_0=5H/g$ while keeping the same values for $\phi_c$ and $\sigma_0$. In this case, the field begins above the crossover scale in most of the ensemble, which prevents the initial decrease in $\langle\phi^2\rangle_{\rm fl}$.

Note that $\langle\phi^2\rangle_{\rm fl}$ does not become exponentially suppressed right after departing from the behavior with $\kappa=0$. Of course, this is because the previous stage of diffusion makes the field's expectation value grow large in some coarse-graining patches. Owing to this, forcing the field below the crossover scale in those patches is a lengthy process indeed. In turn, this implies that the field is able to maintain out-of-equilibrium fluctuations for a longer time in the coarse-graining patches where it has grown larger. As a result of this, there may be a substantial fraction of the ensemble in which the field has a relatively large expectation value compared to the typical equilibrium value.

To better appreciate and quantify this observation we can combine Eqs.~(\ref{eq33}) and (\ref{eq27}) (see Figs.~\ref{fig5} and \ref{fig6}). Consider, for example, the curve $\kappa=10^{-4}gH^2$ in the left-hand panel of Fig.~\ref{fig6}, which corresponds to $\phi_0-\phi_c=0.5H/g$. Using Eq.~(\ref{eq33}) we compute that the field is still fluctuating in 1 to 10\% of the ensemble when $6\times10^2\lesssim g^2N\lesssim4.5\times10^4$ (see Fig.~\ref{fig5}). Introducing this lapse of time in Eq.~(\ref{eq27}) we obtain a value of $\langle\phi^2\rangle_{\rm fl}$ larger than $\phi_c^2$ by a factor of order one only. This implies that $\langle\phi^2\rangle_{\rm fl}$ is never far away from the equilibrium value in a substantial part of the ensemble.

Also, note that the maximum value $g^2\langle\phi^2\rangle_{\rm fl}\simeq38H^2$ is attained at $g^2N\sim5.1\times10^5$, by which time the field is performing out-of-equilibrium fluctuations in $0.2\%$ of the ensemble only. Such a result clearly stems from having set the initial field distribution very close to the crossover scale, which can be avoided by increasing $\phi_0-\phi_c$. We consider such an instance in the righthand panel, for which $\phi_0-\phi_c=2H/g$. From Eq.~(\ref{eq33}) we find that the field performs out-of-equilibrium fluctuations in 1 to 10\% of the ensemble when $8.7\times10^3\lesssim g^2N\lesssim3.6\times10^5$ (see Fig.~\ref{fig5}). Using now Eq.~(\ref{eq27}) in this interval we find a maximum $g^2\langle\phi^2\rangle_{\rm fl}\simeq152H^2$, which is one order of magnitude larger than $\phi_c^2$ (see Fig.~\ref{fig6}) and hence substantially larger than the amplitude of the equilibrium fluctuations.

\section{Conclusions}
In this paper we have investigated the stochastic dynamics of coupled scalar fields during inflation.  We assume a massless scalar field $\Phi$, divide it into short wavelength and long wavelength parts, integrate out the short wavelength modes and obtain stochastic equations of motion for the long wavelength part $\phi$  that we coupled to another (massive) scalar $\chi$ through an interaction of the form $g^2\phi^2\chi^2$. We also assume that the coarse grained patches thus obtained can be treated as separate universes. The problem is generic but has recently been discussed in particular in the context of the flat directions of MSSM  \cite{Enqvist:2011pt,Kawasaki:2012bk}, and $\phi$ could be identified with an MSSM flat direction while $\chi$ would be a non-flat direction.

We allow for the possibility that at the onset of inflation, the massless field $\phi$ has a large expectation value. As a consequence, the scalar $\chi$ becomes a heavy field. This allows us to integrate out the $\chi$ field, thus obtaining an effective dynamics for $\phi$. We find that, despite its coupling, the massless field $\phi$ manages to fluctuate as if free on the Hubble timescale. This happens because the $\chi$ field, while being heavy, adjusts the $\phi$-dependent magnitude of its  fluctuations in a timescale very short compared to the Hubble scale. Owing to this swift response of $\langle\chi^2\rangle$, the random motion of $\phi$ proceeds virtually unimpeded. Thanks to this randomness, the expectation value of $\phi$ grows large in some coarse-graining patches and small in others. Whenever the field's expectation value falls below the crossover scale, the $\chi$ field ceases being heavy and, consequently, the production of its superhorizon fluctuations is no longer suppressed. In turn, the superhorizon fluctuations of the $\chi$ field enhance the interaction potential for $\phi$. We assume that this enhancement traps the field around $\phi=0$ indefinitely, while its fluctuations attain their equilibrium amplitude. This assumption is translated into the field's stochastic dynamics by imposing an absorbing barrier boundary condition for the Fokker-Planck equation located at the relevant scale.

After integrating out the $\chi$ field we solve exactly the effective Fokker-Planck equation for $\phi$, thus obtaining an analytical expression for $P(\phi,t)$. This is done in subsection \ref{sub:FP} and at the beginning of section \ref{sec.fluct}. Using the result, we compute the fraction of the ensemble in which the field's expectation value remains above the crossover scale and the corresponding contribution $\langle\phi^2\rangle_{\rm fl}$ to the mean-square. Our expression for $\langle\phi^2\rangle_{\rm fl}$ reproduces the known result $\langle\phi^2\rangle\propto t$ as long as $\phi$ remains above the crossover scale in most of the ensemble.

However, when $\phi$ falls below the crossover scale in a substantial part of it, the growth of field fluctuations greatly differs from the known result. Instead of growing as $\langle\phi^2\rangle\propto t$, we find that $\langle\phi^2\rangle_{\rm fl}\propto \sqrt{t}$  (see Eq.~(\ref{eq34})).

Although the motion of $\phi$ is fluctuation dominated on the Hubble timescale, for a sufficiently long-lasting phase of inflation and beyond certain timescale the diffusive motion becomes overshadowed by the overall drift towards $\phi=0$ originated by the tilted potential. If the inflation persists beyond such timescale, the probability density for field values above the crossover scale goes to zero exponentially faster than for a purely diffusive motion (see Eq.~(\ref{eq14})). Clearly, this result is due, on one hand, to our assumption on the existence of an absorbing barrier, and on the other hand, to the residual tilt in the effective potential of $\phi$. Therefore, the ultimate fate of the coupled massless field $\phi$ is to become trapped around $\phi=0$ (where the production superhorizon fluctuations of $\chi$ is not suppressed) and fluctuating with the equilibrium amplitude.

As a final comment, we emphasize that although the number of $e$-foldings required to drive the $\phi$ to its equilibrium state is very large indeed, eternal inflation can obviously result in an arbitrary large number of inflationary $e$-folds for our observable universe. If this turns out to be the case, the coupled massless field $\phi$ fluctuates with the equilibrium amplitude when the scales of cosmological interest exit the horizon in the last phase of inflation. However, if inflation does not last  for an exponentially large number of $e$-foldings, the assumed initial expectation value of $\phi$ and the subsequent phase of quasi-free fluctuations can give rise to a pattern of fluctuations in $\phi$ which can be substantially different from the gaussian pattern expected to emerge from equilibrium fluctuations. We illustrate this feature with a particular case in which the amplitude of the field fluctuations $\langle\phi^2\rangle_{\rm fl}$ can become up to ten times larger than the equilibrium amplitude in 1 to 10\% of the ensemble.

\section{Acknowledgements}
JCBS wishes to thank the Helsinki Institute of Physics for hospitality during the course of this research and D. Figueroa for comments and discussions. JCBS is supported by the Spanish Ministry of Science and Innovation through the research projects FIS2006-05895 and  Consolider EPI CSD2010-00064. KE is supported by the Academy of Finland grant 1218322.

\begin{thebiblio}{99}
%\cite{Linde:1982uu}
\bibitem{Linde:1982uu}
  A.~D.~Linde,
  %``Scalar Field Fluctuations in Expanding Universe and the New Inflationary Universe Scenario,''
  Phys.\ Lett.\ B {\bf 116} (1982) 335.

%\cite{Starobinsky:1982ee}
\bibitem{Starobinsky:1982ee}
  A.~A.~Starobinsky,
  %``Dynamics of Phase Transition in the New Inflationary Universe Scenario and Generation of Perturbations,''
  Phys.\ Lett.\ B {\bf 117} (1982) 175.

%\cite{Vilenkin:1982wt}
\bibitem{Vilenkin:1982wt}
  A.~Vilenkin and L.~H.~Ford,
  %``Gravitational Effects upon Cosmological Phase Transitions,''
  Phys.\ Rev.\ D {\bf 26} (1982) 1231.

%\cite{Starobinsky:1986fx}
\bibitem{Starobinsky:1986fx}
  A.~A.~Starobinsky,
  %``Stochastic De Sitter (inflationary) Stage In The Early Universe,''
  in Field Theory, Quantum Gravity and Strings (1986), Vol. 246 of Lecture Notes in Physics, Berlin, Springer, p. 107-126.

%\cite{Affleck:1984fy}
\bibitem{Affleck:1984fy}
  I.~Affleck and M.~Dine,
  %``A New Mechanism for Baryogenesis,''
  Nucl.\ Phys.\ B {\bf 249}, 361 (1985).
  %%CITATION = NUPHA,B249,361;%%

\bibitem{MSSM-REV}
 K.~Enqvist and A.~Mazumdar,
  %``Cosmological consequences of MSSM flat directions,''
  Phys.\ Rept.\  {\bf 380}, 99 (2003).

%\cite{Enqvist:2011pt}
\bibitem{Enqvist:2011pt}
K.~Enqvist, D.~G.~Figueroa and G.~Rigopoulos,
%``Fluctuations along supersymmetric flat directions during Inflation,''
JCAP {\bf 1201} (2012) 053  [arXiv:1109.3024 [astro-ph.CO]].

%\cite{Hosoya:1988yz}
\bibitem{Hosoya:1988yz}
A.~Hosoya, M.~Morikawa and K.~Nakayama,
%``Stochastic Dynamics Of Scalar Field In The Inflationary Universe,''
Int.\ J.\ Mod.\ Phys.\ A {\bf 4} (1989) 2613.

%\cite{Wands:2000dp}
\bibitem{Wands:2000dp}
D.~Wands, K.~A.~Malik, D.~H.~Lyth and A.~R.~Liddle,
%``A New approach to the evolution of cosmological perturbations on large scales,''
Phys.\ Rev.\ D {\bf 62} (2000) 043527  [astro-ph/0003278].

%\cite{Bunch:1978yq}
\bibitem{Bunch:1978yq}
T.~S.~Bunch and P.~C.~W.~Davies,
%``Quantum Field Theory in de Sitter Space: Renormalization by Point Splitting,''
Proc.\ Roy.\ Soc.\ Lond.\ A {\bf 360} (1978) 117.

%\cite{Mijic:1994vv}
\bibitem{Mijic:1994vv}
S.~Habib,
  %``Stochastic inflation: The Quantum phase space approach,''
  Phys.\ Rev.\ D {\bf 46} (1992) 2408  [gr-qc/9208006]; M.~Mijic,
%``Stochastic dynamics of coarse grained quantum fields in the inflationary universe,''
Phys.\ Rev.\ D {\bf 49} (1994) 6434  [gr-qc/9401030].

%\cite{Gardiner}
\bibitem{Gardiner}
C.~W.~Gardiner, ``Handbook of stochastic methods'' (Springer, 1985).

%\cite{Starobinsky:1994bd}
\bibitem{Starobinsky:1994bd}
  A.~A.~Starobinsky and J.~Yokoyama,
  %``Equilibrium state of a selfinteracting scalar field in the De Sitter background,''
  Phys.\ Rev.\ D {\bf 50} (1994) 6357  [astro-ph/9407016].

%\cite{Kawasaki:2012bk}
\bibitem{Kawasaki:2012bk}
M.~Kawasaki and T.~Takesako,
%``Stochastic Approach to Flat Direction during Inflation,''
JCAP {\bf 1208} (2012) 031  [arXiv:1207.1165 [hep-ph]].

%\cite{Dolgov:1982th}
\bibitem{Dolgov:1982th}
  A.~D.~Dolgov and A.~D.~Linde,
  %``Baryon Asymmetry in Inflationary Universe,''
  Phys.\ Lett.\ B {\bf 116} (1982) 329.

%\cite{Abbott:1982hn}
\bibitem{Abbott:1982hn}
  L.~F.~Abbott, E.~Farhi and M.~B.~Wise,
  %``Particle Production in the New Inflationary Cosmology,''
  Phys.\ Lett.\ B {\bf 117} (1982) 29.

%\cite{Kofman:1994rk}
\bibitem{Kofman:1994rk}
  L.~Kofman, A.~D.~Linde and A.~A.~Starobinsky,
  %``Reheating after inflation,''
  Phys.\ Rev.\ Lett.\  {\bf 73} (1994) 3195  [hep-th/9405187]; L.~Kofman, A.~D.~Linde and A.~A.~Starobinsky,
  %``Towards the theory of reheating after inflation,''
  Phys.\ Rev.\ D {\bf 56} (1997) 3258  [hep-ph/9704452].

%\cite{Felder:1998vq}
\bibitem{Felder:1998vq}
  G.~N.~Felder, L.~Kofman and A.~D.~Linde,
  %``Instant preheating,''
  Phys.\ Rev.\ D {\bf 59} (1999) 123523  [hep-ph/9812289].

%\cite{Kofman:2004yc}
\bibitem{Kofman:2004yc}
  L.~Kofman, A.~D.~Linde, X.~Liu, A.~Maloney, L.~McAllister and E.~Silverstein,
  %``Beauty is attractive: Moduli trapping at enhanced symmetry points,''
  JHEP {\bf 0405} (2004) 030  [hep-th/0403001].

%\cite{Watson:2004aq}
\bibitem{Watson:2004aq}
  S.~Watson,
  %``Moduli stabilization with the string Higgs effect,''
  Phys.\ Rev.\ D {\bf 70} (2004) 066005  [hep-th/0404177].

%\cite{Kadota:2003tn}
\bibitem{Kadota:2003tn}
  K.~Kadota and E.~D.~Stewart,
  %``Inflation on moduli space and cosmic perturbations,''
  JHEP {\bf 0312} (2003) 008  [hep-ph/0311240].

%\cite{Bueno Sanchez:2006ah}
\bibitem{Bueno Sanchez:2006ah}
  J.~C.~Bueno Sanchez and K.~Dimopoulos,
  %``Trapped quintessential inflation in the context of flux compactifications,''
  JCAP {\bf 0710} (2007) 002  [hep-th/0606223].

%\cite{Green:2009ds}
\bibitem{Green:2009ds}
  D.~Green, B.~Horn, L.~Senatore and E.~Silverstein,
  %``Trapped Inflation,''
  Phys.\ Rev.\ D {\bf 80} (2009) 063533  [arXiv:0902.1006 [hep-th]].

%\cite{Barnaby:2009mc}
\bibitem{Barnaby:2009mc}
  N.~Barnaby, Z.~Huang, L.~Kofman and D.~Pogosyan,
  %``Cosmological Fluctuations from Infra-Red Cascading During Inflation,''
  Phys.\ Rev.\ D {\bf 80} (2009) 043501  [arXiv:0902.0615 [hep-th]].

%\cite{Wu:2006xp}
\bibitem{Wu:2006xp}
  C.~-H.~Wu, K.~-W.~Ng, W.~Lee, D.~-S.~Lee and Y.~-Y.~Charng,
  %``Quantum noise and a low cosmic microwave background quadrupole,''
  JCAP {\bf 0702} (2007) 006  [astro-ph/0604292].

%\cite{Lee:2011fj}
\bibitem{Lee:2011fj}
  W.~Lee, K.~-W.~Ng, I-C.~Wang and C.~-H.~Wu,
  %``Trapping effects on inflation,''
  Phys.\ Rev.\ D {\bf 84} (2011) 063527  [arXiv:1101.4493 [hep-th]].

%\cite{Vilenkin:1983xp}
\bibitem{Vilenkin:1983xp}
  A.~Vilenkin,
  %``Quantum Fluctuations In The New Inflationary Universe,''
  Nucl.\ Phys.\ B {\bf 226} (1983) 527.

%\cite{Enqvist:1987au}
\bibitem{Enqvist:1987au}
  K.~Enqvist, K.~W.~Ng and K.~A.~Olive,
  %``Scalar Field Fluctuations In The Early Universe,''
  Nucl.\ Phys.\ B {\bf 303} (1988) 713.

%\cite{Riotto:2002yw}
\bibitem{Riotto:2002yw}
  A.~Riotto,
  %``Inflation and the theory of cosmological perturbations,''
  hep-ph/0210162.

%\cite{Berera:2008ar}
\bibitem{Berera:2008ar}
A.~Berera, I.~G.~Moss and R.~O.~Ramos,
%``Warm Inflation and its Microphysical Basis,''
Rept.\ Prog.\ Phys.\  {\bf 72} (2009) 026901  [arXiv:0808.1855 [hep-ph]].

%\cite{Kamada:2009hy}
\bibitem{Kamada:2009hy}
  K.~Kamada and J.~Yokoyama,
  ``On the realization of the MSSM inflation,''  Prog.\ Theor.\ Phys.\  {\bf 122} (2010) 969  [arXiv:0906.3402 [hep-ph]]; K.~Kamada and J.~'I.~Yokoyama,
  ``Dissipative effects on MSSM inflation,''  Int.\ J.\ Mod.\ Phys.\ Conf.\ Ser.\  {\bf 01} (2011) 114.

%\cite{BasteroGil:2006vr}
\bibitem{BasteroGil:2006vr}
  M.~Bastero-Gil and A.~Berera,
  %``Warm inflation dynamics in the low temperature regime,''
  Phys.\ Rev.\ D {\bf 76} (2007) 043515  [hep-ph/0610343].

%cite{Risken}
\bibitem{Risken}
H.~Risken, %The Fokker–Planck Equation: Method of Solution and Applications,
Springer-Verlag, New York, 1989.

%\cite{Enqvist:2012xn}
\bibitem{Enqvist:2012xn}
  K.~Enqvist, R.~N.~Lerner, O.~Taanila and A.~Tranberg,
  %``Spectator field dynamics in de Sitter and curvaton initial conditions,''
  arXiv:1205.5446 [astro-ph.CO].

%\cite{Linde:1986fc}
\bibitem{Linde:1986fc}
A.~D.~Linde,
%``Eternal Chaotic Inflation,''
Mod.\ Phys.\ Lett.\ A {\bf 1} (1986) 81.

%\cite{Chandrasekhar:1943ws}
\bibitem{Chandrasekhar:1943ws}
S.~Chandrasekhar,  %``Stochastic problems in physics and astronomy,''
Rev.\ Mod.\ Phys.\  {\bf 15} (1943) 1.

%cite{Cox}
\bibitem{Cox}
D.R. Cox, H.D. Miller, %The Theory of Stochastic Processes,
Chapman \& Hall, CRC, Boca Raton, Florida, USA, 1965.

%cite{Molini}
\bibitem{Molini}
A.~Molini, P.~Talkner, G.~G.~Katul, A.~Porporato, %First passage time statistics of Brownian motion with purely time dependent drift and diffusion
Physica A 390 (2011) 1841-1852.

%\cite{Lorenz:2010vf}
\bibitem{Lorenz:2010vf}
  L.~Lorenz, J.~Martin and J.~'i.~Yokoyama,
  %``Geometrically Consistent Approach to Stochastic DBI Inflation,''
  Phys.\ Rev.\ D {\bf 82} (2010) 023515  [arXiv:1004.3734 [hep-th]].

\end{thebiblio}
\end{document}